# A Retrieval Framework and Implementation for Electronic Documents with Similar Layouts

*Hyunji Chung*

Korea University, Republic of Korea

## Abstract

As the number of digital documents requiring investigation increases, it has become more important to identify relevant documents to a given case. For instance, in e-discovery, processing electronic document files is one of the most significant tasks because attorneys usually have to review large amounts of documents and select specific ones related to cases. By the way, if people are skilled at revising documents like touching contents, replacing words and changing the language, identifying relevant documents will be more time consuming and less accurate. Hence, there have been continual demands for finding relevant (or similar) files in order to overcome this kind of issues.

Regarding finding similar (possibly relevant) files, there can be a situation where there is no available metadata such as timestamp, filesize, title, subject, template, author, etc. In this situation, investigators will focus on searching document files having specific keywords related to a given case. Although the traditional keyword search with elaborate regular expressions is useful for digital forensics, there is a possibility that closely related documents are missing because they have totally different body contents.

In this paper, we introduce a recent actual case on handling large amounts of document files. This case suggests that 'similar layout' search will be useful for more efficient digital investigations if it can be utilized appropriately for supplementing results of the traditional keyword search. Until now, research involving electronic-document similarity has mainly focused on byte streams, format structures and body contents. However, there has been little research on the similarity of visual layouts from the viewpoint of digital forensics. In order to narrow this gap, this study demonstrates a novel framework for retrieving electronic document files having similar layouts, and implements a tool (SSDOC) for finding similar Microsoft OOXML files using user-controlled layout queries based on the framework.

## 1 Introduction

Currently, as the number of digital devices requiring investigation increases and digital document formats become increasingly complicated, investigators spend significant time examining digital documents [1, 2]. As long as there are enough trained investigators to meet the proliferation of digital devices, it may be possible to analyze data manually. However, realistically, increasing the number of investigators cannot match the increasing rate of data volume and complexity [3]. In this situation, automatically filtering similar (possibly relevant) files can save time and increases the accuracy of digital investigations [4]. Moreover, techniques for assessing similarity are considered to be an essential tool for advanced forensic analysis [5]. For these reasons, the stakeholders of digital forensics are requiring more powerful techniques addressing similarities in digital documents.

Specifically, techniques used to determine electronic-document similarity are useful for the process of e-discovery. During the e-discovery process, certain digital documents related to a given case are usually selected from among enormous amounts of electronic documents [6]. In such a case, it is imperative to automatically filter electronic documents with regard to a given case in order to save time [6]. It is also very expensive for clients to engage attorneys in the e-discovery process based on the time spent considering the relevance of each document.

One approach to reducing such costs is the use of appropriate document classification and information retrieval [6]. Until now, keyword-based searches and classification methods have been used in general [7]. Futhermore, existing studies have focused on the similarity of byte streams, file formats, and body text [9–27]. Although all of them are meaningful for digital forensic activities, it is necessary to consider various aspects of electronic-document similarity since using only these techniques is not enough for more complex circumstances.

In digital investigations, there can be a situation where it is necessary to identify relevant document files in connection with a crime. Possible relevant files may have similar contents, and so they could be found by analyzing body text and metadata (e.g., timestamp, author, last saved by) in general. However, the traditional keyword search would be both costly and time-consuming if there are too many results that should be manually reviewed [8]. In addition, it would be more difficult to find relevant files if there is no available metadata such as timestamp, filesize, title, subject, template, author, etc. Besides, existing approximate matching algorithms based on byte streams and contents will also fail if some potential relevant files have totally different contents.

Thus, this study focuses on visual layouts[1] as a novel concept for enhancing the existing similar document file search. It is important note that the layout of an electronic document discussed here does not mean just applying templates or themes provided by document editing applications. That is to say, although document files use a specific template and theme, each layout entity such as, for instance, body text, images and tables can be placed everywhere in a page (slide or sheet) depending on user contents. Regarding this study, we will introduce a recent actual case in the next section to emphasize our motivation because there has been little research on the similarity of visual layouts from the viewpoint of digital forensics.

In this paper, we propose a new method for retrieval of electronic documents having similar visual layouts. The following summarizes three contributions:

- This work suggests a concept of layout similarity, and demonstrates the significance and necessity of the similar layout search through a recent actual case.
- This work proposes a new framework for retrieving document files having similar layouts.
- This work introduces a tool, SSDOC that is implemented for finding similar Microsoft OOXML files using user-controlled layout queries based on the proposed framework.

This paper is organized as follows: Section 2 explains a more detailed motivation of this paper with an actual case, and Section 3 summarizes related works. Section 4 proposes a novel framework for retrieving digital documents that potentially include similar layouts. Section 5 introduces SSDOC, a tool that is implemented based on the framework proposed in Section 4. Section 6 performs an experiment for verifying and evaluating the proposed framework using a public dataset. Finally, Section 7 presents our conclusions.

[1] According to diplomatics that is a scholarly discipline centered on the critical analysis of documents, the layout of a paper document describes rectangular frames that include content [28]. Similarly, a layout in a digital document means an arrangement of text or graphics [29] that includes various properties associated with, for example, text boxes, images, tables, shapes and fonts.

# 2 Motivation

## 2.1 Problem Definition

This study started to address issues raised by a recent actual case in South Korea. In this case, there was a situation where examiners suspected some government employees worked for an election campaign. For reference, Public Official Election Act in South Korea bars government employees from running an election campaign. The following summarise important details with regard to our study motivation. Figure 1 will help you understand the overall meaning.

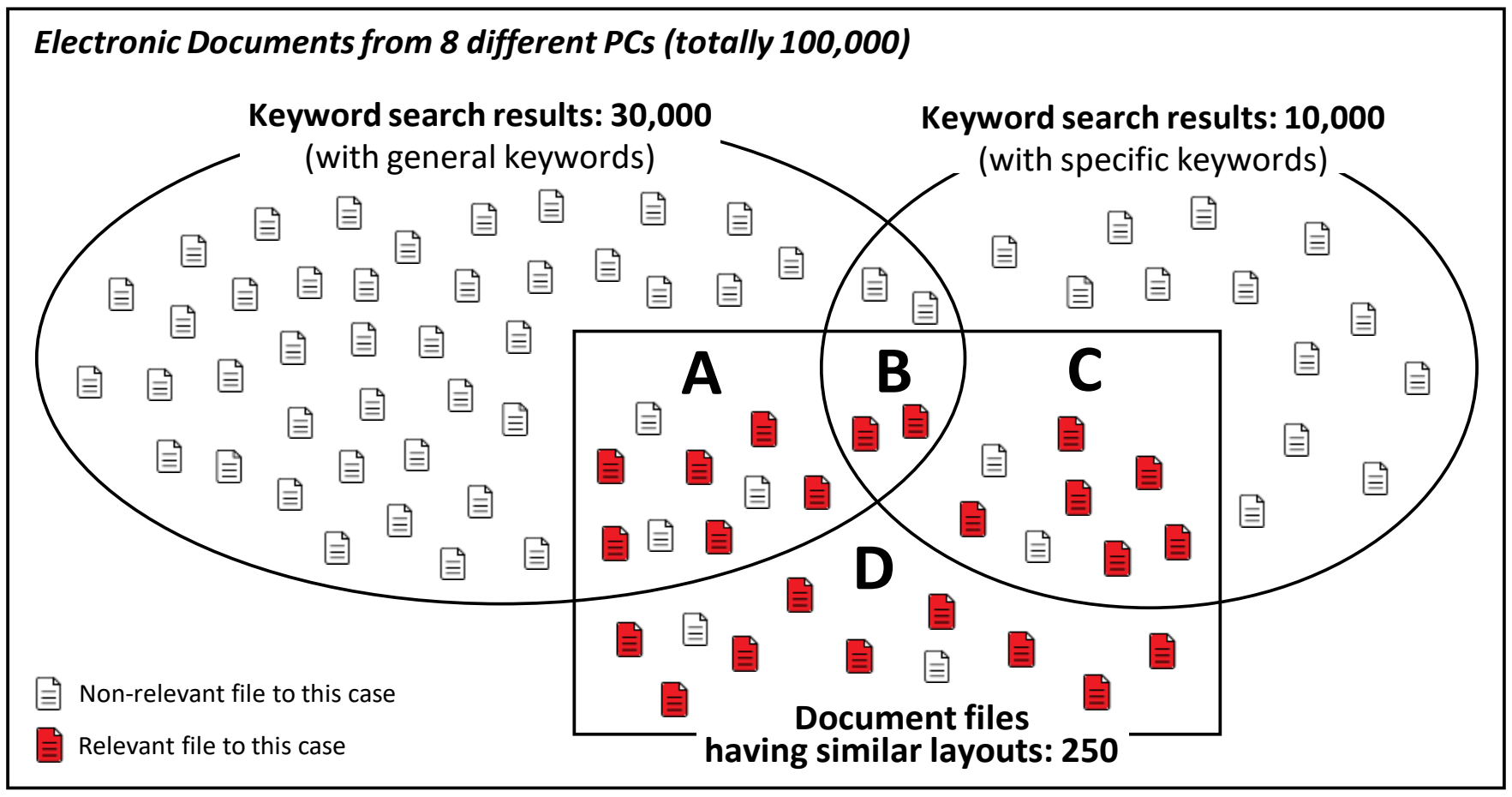

*Figure 1. Motivation of this study: 'similar layout' search may be useful for digital investigation if it can be utilized appropriately for supplementing results of the traditional keyword search*

One day, the examination team of an election commission received interesting information from an anonymous informant. This information was about government employees worked for supporting and promoting a front running candidate 'X'. In detail, the informant got an email with a document file (named 'INFO') on supporting candidate 'X' from the communication team of a district office. After a while, digital forensic examiners performed an objective and exhaustive analysis on PCs related to the suspicious communication team of a district office. After finishing the initial analysis and recovering deleted files (by data carving), there were approximately 100,000 document files having the same format as 'INFO' file. In particular, they wanted to find any document files which have relevance to 'INFO' file received from the informant.

For that, when they first tried to do the keyword search using general words from the body text of 'INFO' file such as exact names of political parties and candidates, they got approximately 30,000 document files as a result of the first search. Also, when searching files using more specific words related to candidate 'X', they found about 10,000 document files. In order to classify relevant documents, examiners had to manually review all files detected by the keyword search. As a result of it, they found 100 document files that seemed to be relevant to the promotion of candidate 'X' (See area A, B and C in Figure 1). An interesting fact was that the files not used an exactly same template, but had very similary layouts related to page size (width, height), text properties (size, font, color), image properties (size, position), table properties (row/column, size, position), etc. Through additional examinations of suspects, it was revealed that the files were written as weekly reports for the last few years. With this findings, examiners decided to analyze whole document files except for the one found via the keyword search because of the possibility of non-specified clues on violating the election law.

After going through the exhaustive analysis, additional 100 document files were found. Although they used very similar layouts with previously found document files, their contents were totally different because they were created for supporting candidate 'Y' (See area D in Figure 1). Consequently, 8 government employees were charged with violating Public Official Election Act because they supported two candidates using more than 200 document files.

In the above case, it was hard to filter by timestamp because there was no a specific period of time for finding relevant files. It was also a situation where exaiminers could not utilize metadata such as title, subject and author since it was deleted by a security policy. Thus, without the additional analysis on whole files, exaiminers would not have been able to find document files relate to candidate 'Y'.

In summary, this study started with a question about how to find relevant document files by using the similarity of visual layouts. In this paper, we propose a novel framework for the layout retrieval motivated by an idea that the 'similar layout' search may be useful for more efficient digital investigations if it can be utilized appropriately for supplementing results of the traditional keyword search. It should be noted that this work focuses not on the keyword search but on the similar layout search.

## 2.2 Electronic Documents and Data Similarity

In order to address the issues presented in Section 2.1, we reinterpret and explain data similarity from the viewpoint of electronic documents.

Table 1 summarizes similarity types that exist in an electronic document. First, algorithms exist for calculating raw data-based similarity. Raw data-based similarity focuses on calculating the similarity of byte streams. In digital forensics, typical algorithms used to determine byte-stream-based similarity include ssdeep, sdhash, MRSH-v2, TLSH, etc. [8–11]. Second, structure-based similarity describes similarity of internal structures in which a digital document saves data. In this context, most research focuses on XML, which is widely used for saving and exchanging digital data. Third, layout-based similarity describes similarity of visual layouts, such as text box position and size, table position and size, image position, or the page (or slide or sheet) size of a digital document. There are some researches about layout similarity. Finally, content-based similarity describes similarities in the content of digital documents. There are many studies that focus on measuring content similarity based on keywords and semantic analysis results of body text.

*Table 1. Similarity types relating to electronic documents*

| Similarity Type | Description |
|---|---|
| Raw data based similarity | • Similarity of byte streams, regardless of analysis of the internal file format and content<br>• ex) sdhash, ssdeep, MRSH-v2, TLSH |
| Structure based similarity | • Similarity of the internal format structure containing information related to the electronic document<br>• ex) tree structure of XML/HTML format |
| Visual layout based similarity | • Similarity of visual layouts<br>• ex) image properties (position, size), table properties (position, size), cell properties (pattern, color) |
| Content based similarity | • Similarity of body text<br>• ex) keyword-based analysis, concept-based analysis, text stylometry |

## 3 Related Works

There are various studies focusing on similarity associated with PDF, HTML, XML, OOXML (XML level), and spreadsheets. Based on the similarity types defined in Section 2.2, Table 2 shows the summary of previous works. Note that raw data-based similarity algorithms, such as ssdeep, sdhash, MRSH-v2, TLSH, are excluded from this summary because they are widely known in the digital forensics community.

*Table 2. Summary of previous works*
*(Meaning of symbols in the 2nd column: S = Structure, L = Layout, C = Content)*

| Previous work | Similarity | | | Document type | Method to measure similarity |
|---|---|---|---|---|---|
| | S | L | C | | |
| B. Rosenfeld et al. (2002) [13] | - | O | - | PDF | - Graph mapping algorithm |
| I. F. Cruz et al. (2006) [14] | O | - | O | HTML | - Tag frequency distribution analysis<br>- Parametric function<br>- Edit distance |
| S. Flesca et al. (2002) [15] | O | - | - | XML | - Fast fourier transform |
| A. Nierman et al. (2002) [16] | O | - | - | XML | - Tree edit distance |
| W. Liang et al. (2005) [17] | O | - | - | XML | - LAX join algorithm |
| J. Tekli et al. (2006) [18] | O | - | O | XML | - Semantic similarity<br>- Edit distance |
| J. Tekli et al. (2007) [19] | O | - | - | XML | - Tree edit distance |
| W. Kim (2008) [20] | O | - | O | XML | - EDFS(Extended Depth First Search) string match<br>- Content tree's node similarity |
| Y. Watanabe et al. (2012) [21] | O | - | - | OOXML (XML level) | - Advanced LAX join algorithm |
| A. Auvattanasombat et al. (2013) [22] | O | - | O | OOXML (XML level) | - LAX join algorithm + Keyword based similarity |
| A. Auvattanasombat et al. (2013) [23] | O | - | O | OOXML (XML level) | - Keyword based advanced LAX join algorithm<br>- LAX join algorithm + Keyword based similarity |
| S. Chatvichienchai (2011) [24] | - | O | O | Spreadsheet | - Similarity of layouts in body contents |
| S. Chatvichienchai (2013) [25] | - | O | O | Spreadsheet | - Similarity of layouts in body contents |
| F. Cesarini et al. (2002) [26] | - | O | - | Document image | - Similarity of global features<br>- Similarity of occurrences of tree patterns |
| L. Liu et al. (2013) [27] | - | O | - | Document image | - Similarity of graphs generated from document images |

B. Rosenfeld et al. proposed a method that represented visual layouts of Acrobat PDFs as graphs in order to identify digital documents having similar layouts by using the graph-mapping technique [13]. It was meaningful because it studied a method of determining digital document layout similarity.

I. F. Cruz et al. studied the structural and content similarity of HTML tags [14]. Structural similarity was analyzed by calculating the frequency of each tag in HTML based on the frequency distribution analysis. The authors also proposed a method that formulated the structure of HTML tags and measured similarity using the distance between each function.

S. Flesca et al. proposed a method that used Fourier transforms to measure the structural similarity of XML files. Specifically, they proposed to represent XML documents as time series, and computed the structural similarity between two files by using the discrete Fourier transform of the corresponding signals [15].

A. Nierman et al. proposed a method that computed the structural similarity using tree-edit distances between XML documents and classified them using distance values [16]. They found that the clustering results matched the original document type definitions (DTDs). This research demonstrated performance superior to methods previously used for measuring tree similarity.

W. Liang et al. proposed LAX (leaf-clustering-based approximate XML-join algorithm), which computed the structural similarity between XML files. LAX is an algorithm that separates XML documents into subtrees, and calculates similarity between them by determining the similarity degree based on the leaf nodes of each pair of subtrees [17]. However, this method has limitations. Even though contents of subtrees are similar, the structures of subtrees can be substantially different. In this case, although digital documents have similar contents, their similarity scores could be low.

J. Tekli et al. proposed a method for identifying the structural and semantic similarity between documents using a combination of the edit-distance and semantic-similarity algorithm [18]. They also proposed an algorithm for measuring structural similarity between XML and DTD using the tree-edit distance [19].

W. Kim proposed a new method to measure similarity between XML documents by considering their structures and contents. Using the proposed method, documents that were structurally identical or contained similar internal structures were efficiently identified using a string-matching technique [20].

Y. Watanabe et al. analyzed the similarity of XML fragments in digital documents. Similar to previous work [17], this paper proposed LAX+, which was an advanced LAX algorithm for comparison of XML tree structure. This algorithm compared leaf nodes of document.xml (MS Word DOCX file), worksheetN.xml (MS Excel XLSX file), and slideN.xml (MS PowerPoint PPTX file) in compressed XML fragments from an OOXML file [21].

A. Auvattanasombat et al. proposed the KLAX algorithm, which was an advanced version of LAX+ that considered the contents of XML leaf nodes. KLAX calculates LAX+, including keyword similarity. They also proposed the LAX&KEY algorithm, which calculates LAX+ similarity and keyword similarity separately, then combines them [22, 23].

S. Chatvichienchai suggested a method for searching similar spreadsheet documents, such as MS Excel and Lotus1-2-3. He structuralized the contents of documents, and then compared structures of the contents. This research was meaningful because the author utilized the structures of the document content. However, if some parts of the content or visual layout changed, it would not perform appropriately [24, 25].

F. Cesarini et al. proposed a method for retrieval based on the layout similarity of document image files. Pages of document image files were represented with global features and features related to the MXY tree layout. The similarity was computed by combining the similarity measures that were defined for both features [26].

L. Liu et al. proposed an approach that aimed to match near-duplicate document image files using a graphical perspective. In this study, a graph was used to represent a document image, and the nodes in the graph corresponded to the objects in the image, while the edges described their relationships. That is, the document image-matching problem was converted to a graph-matching problem. Using two graphs, the similarity between them was computed [27].

As shown in Table 2, previous studies mainly attempted to measure structural or content similarity at the XML level. There are also some studies concerning retrieval of document image files having similar layouts. However, this method is appropriate for scanned or picture document files only. In the case of electronic documents, these methods are not appropriate for retrieval on the basis of layout similarity for two reasons: 1) if document image retrieval methods are used for electronic documents, it can be inefficient and inaccurate, because electronic documents have values related to visual layouts in their internal format; if these values are used for retrieval, it will be more efficient and accurate; 2) it is difficult for users to control queries for retrieval, because one entire page of a document image is used as query data, and features are extracted automatically, as in the case of document image retrieval. For example, suppose that there is one slide in a presentation file and it includes a table, text balloon, image, and a text box. An investigator wants to search slides that have a 3 × 4 table in the upper left and an image (width: 5 cm, height: 9 cm) in the lower right. In this case, it is useful for users to allow manual input of queries. This is why a new method is needed in order to search digital documents having similar layouts.

# 4 E-Document Layout Retrieval Framework

## 4.1 Framework Overview

Figure 2 is overall concept of framework for retrieval of electronic documents that include similar layouts. The upper panels of Figure 2 relate to the extraction of layout features from all files of target datasets, and the lower panels concern layout retrieval using user queries. Although this framework can be applied to all types of digital documents such as wordprocessing documents, presentations, spreadsheets and drawings, we will mainly explain the overall processes using document files saved in Microsoft OOXML format which is one of the most widely utilized document format standards. It is important to note again that this framework can be generalized to other document formats except OOXML formats if we can interpret their internal structures and extract virsual layout features.

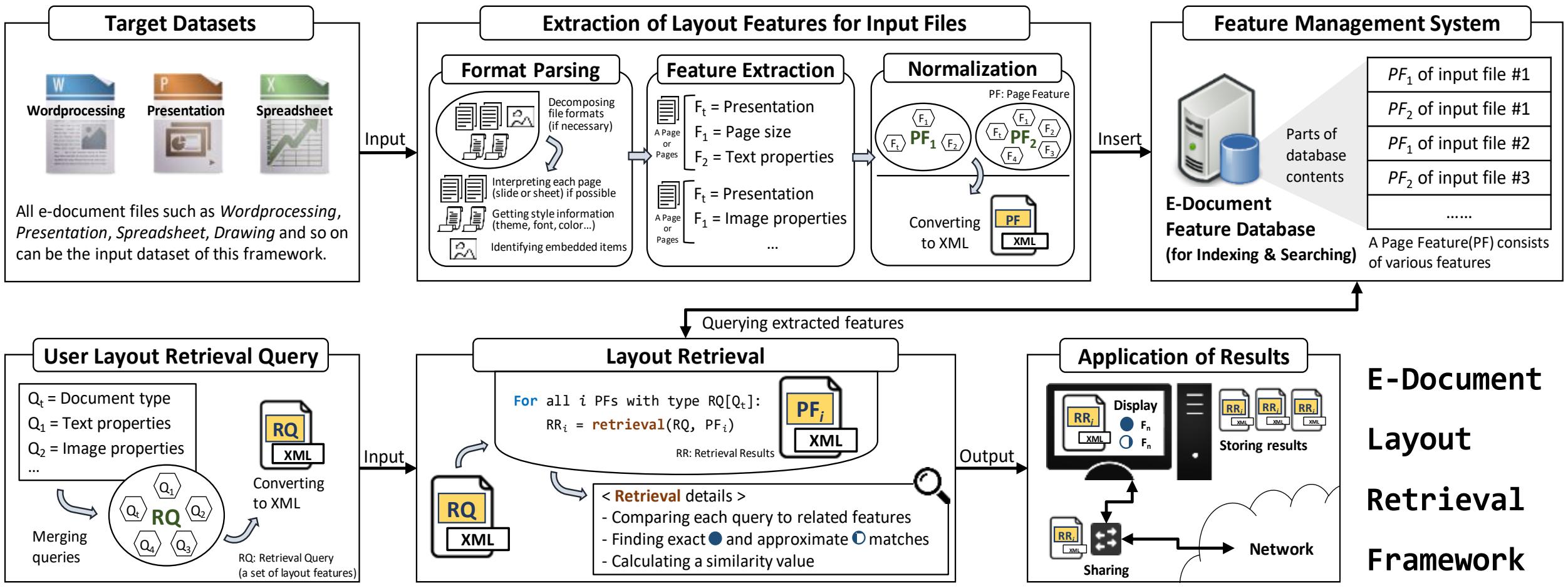

*Figure 2. Overall concept of e-document layout retrieval framework*

## 4.2 Extraction of Layout Features

Layout features are extracted from each page (slide or sheet) of input target datasets. There are three steps involved in the extraction of layout features from OOXML files.

The first step involves parsing the file format of OOXML files. In this stage, each page (slide or sheet) of the file is interpreted in detail. The interpretation process is based on embedded items and style information, such as table size, image size, font name, or font color. In some cases, data is stored with archiving. In this case, the file format of electronic documents needs to be decomposed or decoded in order to acquire valid data from the container internal format.

The second step is to extract layout features from parsed datasets. Layout features constitute visual information that can be seen with the naked eye. For example, OOXML presentation’s layout features are page size (width, height, margins), text properties (size, font, color), image properties (size, position), cell (color, border), table properties (row/column, size, position) and shape (type, size, position). If one or more slide exists in the digital document, the features of all slides are extracted to allow for thorough retrieval. In this study, objects that are widely used in OOXML files (DOCX, PPTX, XLSX) related to slide size, text boxes, images, tables, cells and predefined shapes are chosen as layout features for developing the prototype tool in Section 5. Apart from these features, there are other various types of objects, including diagrams, OLE objects, charts, etc., that can also added in the future.

Finally, the third step involves the normalization process in order to save, manage, and apply the extracted features. Specifically, extracted features from each page (slide or sheet) of a file are converted into XML (or JSON) format and saved in a feature database. A feature management DBMS can be either an independent high-performance system or a small standalone database that is used for layout retrieval later.

More detailed processes with an example Presentation file are outlined in Figure 3. The ‘OOXML presentation format’ box in Figure 3 represents the internal file format in an OOXML presentation file from the target dataset that was used. As shown in the figure, an OOXML presentation file consists of multiple XML files: slideLayout#.xml, slideMaster#.xml, and theme#.xml are used for parsing a slide#.xml file, which has the content and visual layouts of the slide. According to the OOXML standard, they are linked by relationship files in “_rels” folders. Using these files together, a slide#.xml file can be interpreted completely. In the format of slide#.xml, it includes information about text boxes, images, tables, and shapes. Afterward, features extracted from slide#.xml are stored as normalized forms to compare with user queries.

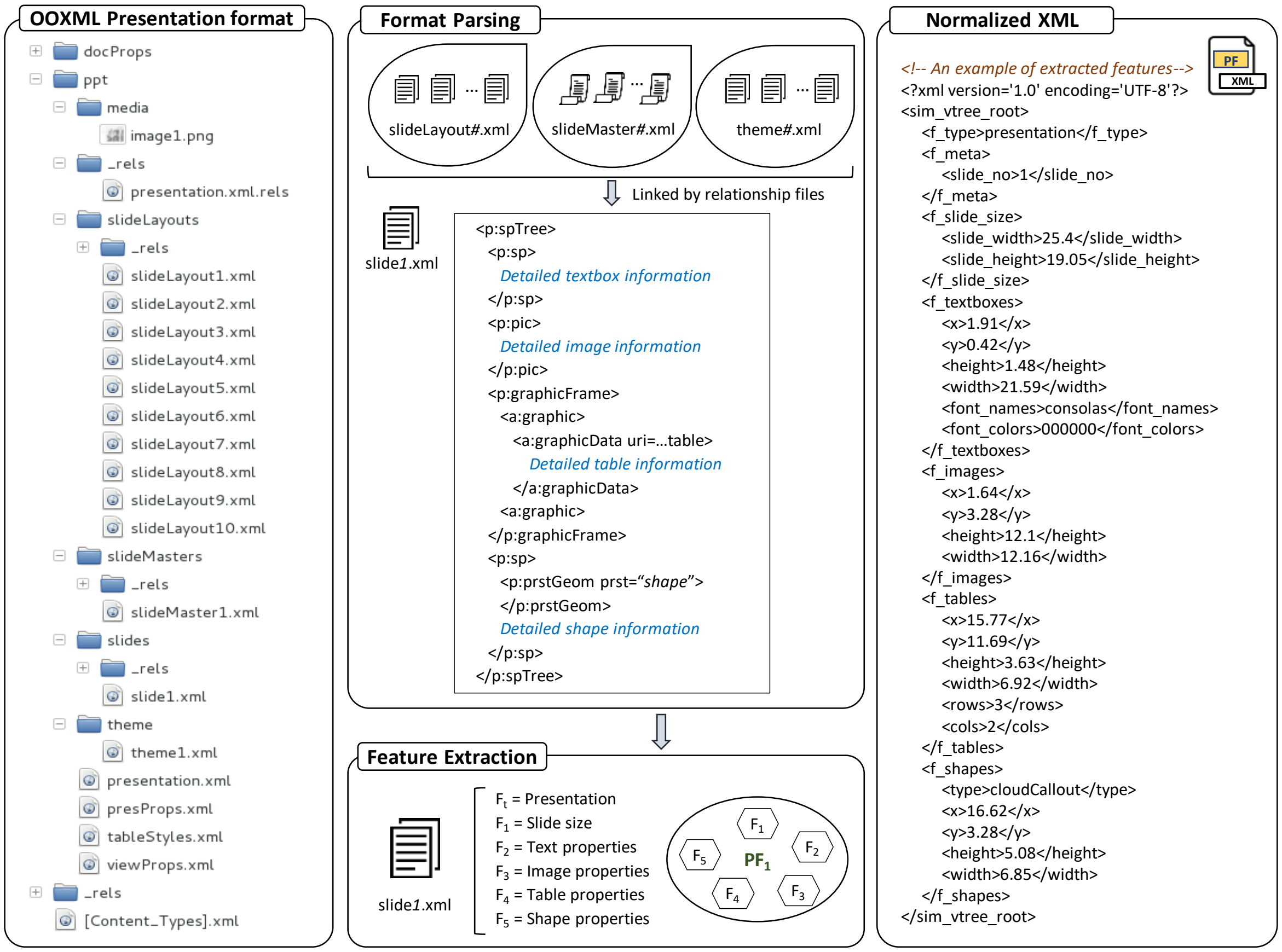

*Figure 3. Extraction processes of layout features from a sample OOXML Presentation file*

Similary, it is possible to extract features from Wordprocessing and Spreadsheet formats. A Wordprocessing file defined by the OOXML format is composed of multiple XML files which have relationships with each other: header#.xml, footer#.xml, endnotes.xml, footnotes.xml, styles.xml, settings.xml, and theme#.xml are utilized for interpreting a document.xml file, which contains the content and visual layouts of the document. There are body text, endnotes, footnotes, headers, footers, TOCs (table of contents), images, and tables as representative layout features. In addition, we can also extract layout features from a Spreadsheet file defined by the OOXML format using the same concepts described above. In this format, sheet#.xml files which have the body content of each worksheet includes various information about cells (each cell has a style including font, fill and border), images, and charts. These features can be acquired by parsing sheet#.xml files together with related files such as drawing#.xml, chart#.xml, workbook.xml, and styles.xml.

## 4.3 Layout Retrieval Details

### 4.3.1 Building User Layout Retrieval Queries (RQs)

Users are required to enter queries for layout retrieval. The data types of user queries are divided into two parts: numeric and character. The numeric type consists of doubles, integers, and hexadecimal types. For example, height, width and coordinates are double types, the number of rows and columns are integer types, and font color is a hexadecimal type. In the case of integer and double types, users should configure basic units, such as centimeters or inches, in order to process values correctly. Appendix B shows detailed query types and samples of building user queries. Afterwards, user queries are changed to XML (or JSON) format, and a retrieval query (RQ) is created for the next step. For your guidance, in Microsoft Office applications, 'Format shape dialog box' will be useful for identifying detailed values relating to each layout object such as a text box, image, table, cell, or shape.

### 4.3.2 Layout Retrieval Algorithms

The process of finding documents having similar layouts with user queries is performed. If a pre-generated database that manages layout features exists, a more efficient retrieval using the database can be performed on a target of the page feature (PF) that is the same as the document type ($Q_t$) associated with the user query. PF (Page feature) in Figure 2 is symbolic term that describes page features (in the case of a wordprocessing document), slide features (in the case of a presentation), or sheet features (in the case of a spreadsheet).

The retrieval algorithm uses two methods for retrieving matched objects: exact matching (EM) and approximate matching (AM). EM is a method of searching for perfectly matching pages with queries. If all of queries are exact matches, the similarity value (ranges from 0 to 1) is 1. Two pages look similar by the naked eye, however, layout feature values in an internal file format can be slightly different. This is why we also propose four different AM methods of calculating the similarity value according to query types (see Appendix A). In the AM stage, the similarity between a user query and the target object is calculated. The less the similarity, the closer the similarity value will be to 0. It is important to note that additional AM methods for various data types can be added through further studies of course.

In cases of AM-1, the rate of exactly matching features and queries is calculated as the sum of results of EM divided by the number of queries. In cases of AM-2, if the input queries are the same as extracted features, the similarity value is 1. If not, the similarity value is 0. If the type is the same, but the dimension is different, the similarity value is 0.5. For example, when the chart type of a query describes a 2-dimesional bar chart and the chart type of a feature from a spreadsheet document is a 3-dimensional bar chart, the similarity value would be 0.5. In cases of AM-3, the possible range of maximum distances is calculated between the query and the feature, and similar values are calculated by exponential distance. For instance, suppose that the possible range of slide height in a presentation document is from 2.54 to 142.24 cm. When a user query represents slide height as 19.05, the distance range within arbitrary documents would be 0~123.19 (=142.24−19.05). Therefore, the maximum gap is 123.19. The reason we use exponential distribution is that applying an exponential distribution is more precise than a linear distribution empirically for determining wheter two different pages look similar or not by the naked eye. Finally, AM-4 is similar to AM-3. The only difference is that AM-4 is used when a data type involves a coordinate (x, y). For instance, suppose there exists an image in a slide of presentation, where the width of the slide is 25.4, the height of the slide is 19.04, and the query about the position of the image is (3.25, 4.22). The distance range is $0 \sim \sqrt{(25.4-3.25)^2+(19.04-4.22)^2}$. Appendix B is a table that summaries the methods used to measure according to layout types, which are based on Appendix A. Using the AM algorithm, a similarity value ($s_value_{final}$) for each page (slide or sheet) is calculated by adding all similarity values ($s_value_i$) for all user queries as the following:

$$s_value_{final} = \frac{1}{n}\sum_{i=1}^{n} s_value_i$$ , where *n* is the number of queries.

### 4.3.3 Application of Layout Retrieval Results

The retrieval results (RR) have various applications. After RRs are saved in the management system, they can be utilized for advanced information retrieval or analysis activities. Additionally, sharing for co-work or distributed processing is possible through a network, given that RRs are saved in normalized XML (or JSON) format.

# 5 Implementation

## 5.1 Overview of SSDOC (Similarity Search for e-DOCuments)

This Section introduces SSDOC which is a prototype program based on the framework described in Section 4. SSDOC is implemented with Python 3.4 and QT 5.4 as a programming language. The framework described in Section 4 is not limited only to specific document types, enabling all types of documents to be potential targets. In this paper, the prototype program is implemented in order to verify and evaluate the suggested framework. Note that the current version of SSDOC (1.0) is subject to presentation format (PPTX), spreadsheet format (XLSX) and wordprocessing format (DOCX) based on OOXML (Office Open XML). SSDOC v1.0 will extend to support various types of electronic documents in the future.

SSDOC is a freeware tool that can be downloaded from the following URL[2]. The current version of SSDOC does not use additional third party modules for interpreting the OOXML format or extracting layout features. We attempted to implement the prototype code simply and clearly with only a ZIP file and a XML handler for overcoming problems mentioned in Section 2.1.

The experimental prototype tool attempts to access extracted features stored in memory directly instead of storing and querying them using a database system. If this part is implemented with an efficient database system based on the framework described above, it can be applied to process and index large-scale datasets.

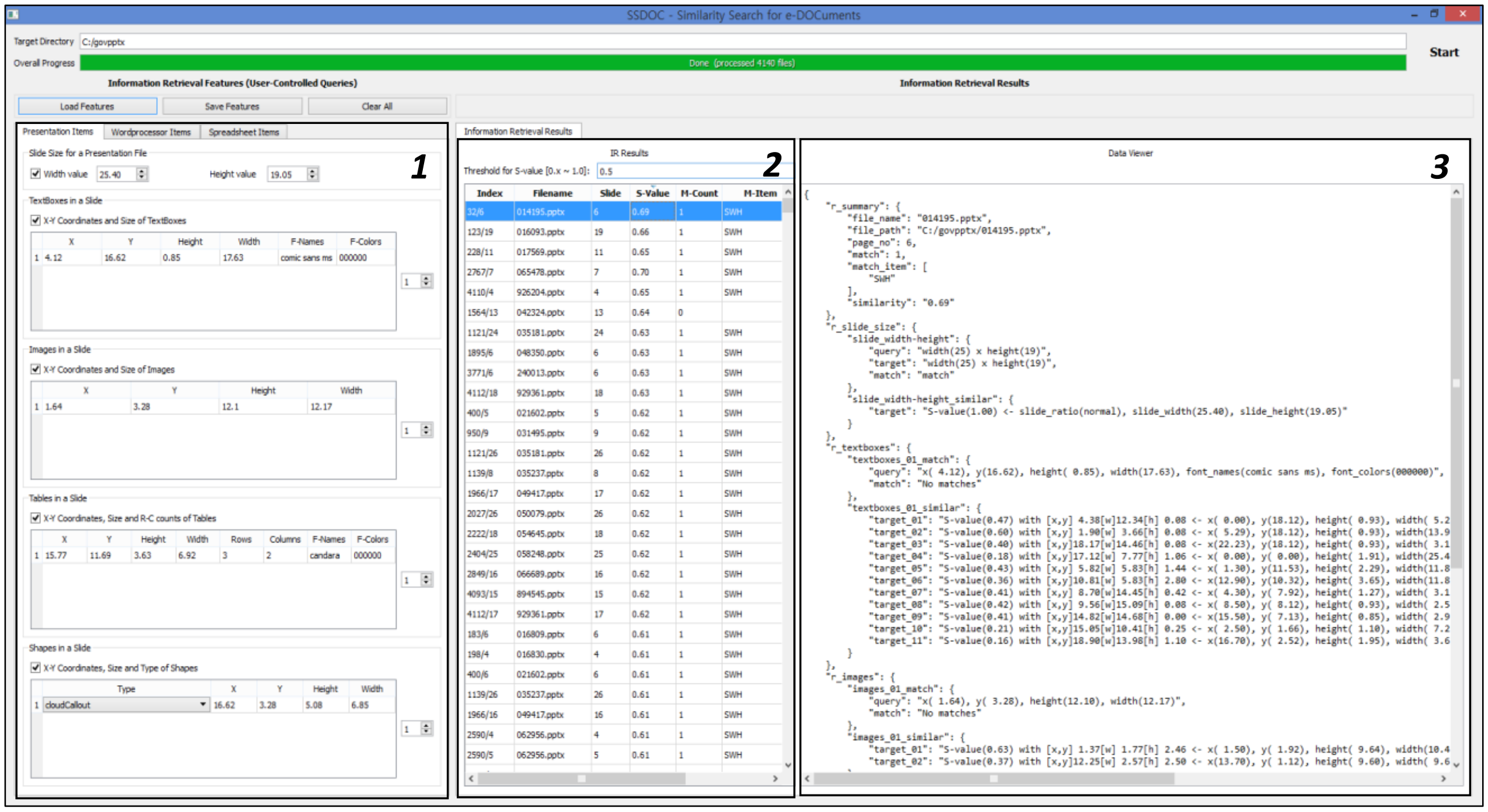

*Figure 4. User interface of SSDOC*

[2] https://sites.google.com/site/datasimilarity/dataset

## 5.2 Class Diagram of SSDOC

Figure 5 represents the class diagram of SSDOC version 1.0, consisting of five modules. One of the main classes is SSDocMainDialog, which provides the user interface. The input for the class is a directory of target datasets and user queries. The directory of target datasets and user queries are passed to the SSDocCore class. For communication between the user interface and the SSDocCore class, WorkerThread is implemented. For discriminating file types, SSDocCore traverses all files in the target directory. When the type of the target file is the same as the type of query, a FileParser module is created and begins searching. For example, if the type of file is PPTX and the user query is related to PPTX, FileParserPPTX is created and begins searching. The FileParser module can be classified by the file format type. In this prototype, FileParsers for PPTX, DOCX and XLSX consist of various members, such as those shown in Figure 5. FileParsers parse the internal structure of the container format and extract the layout features of each page (slide or sheet). Then, comparing the extracted features with user queries, layout similarity is calculated. The extracted features and retrieval results are managed using a python dictionary, JSON, and XML structures using the SimVtree class.

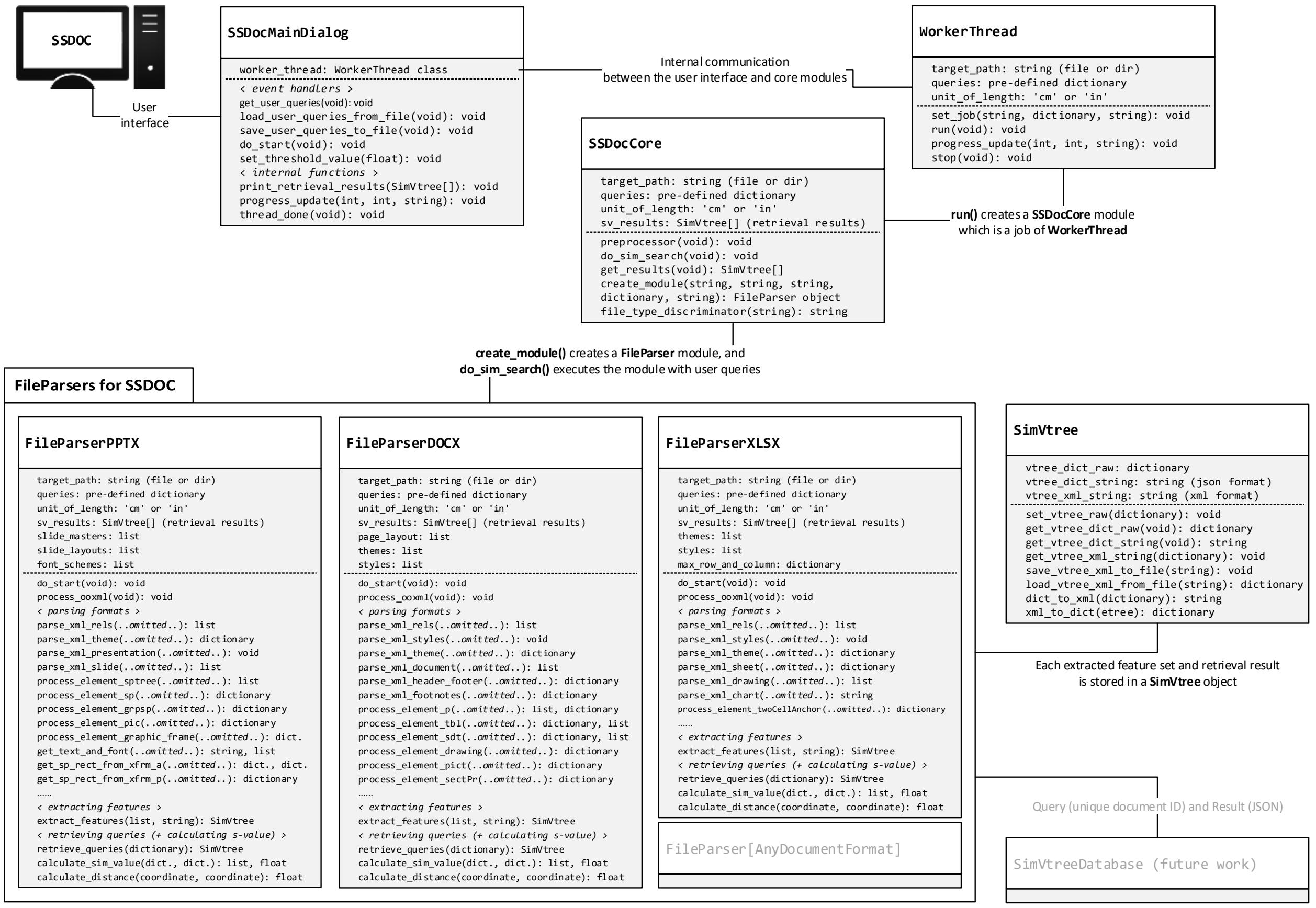

*Figure 5. Class diagram of SSDOC v1.0*

## 5.3 Testing and Evaluation

The GOVDOCS corpus is used as a test dataset for verifying and evaluating the developed prototype tool[3]. Since MS Office document files from the corpus are almost all Microsoft Compound File Binary Format (CFBF) files, we attempted to convert CFBF files to OOXML files using Office File Converter (OFC) [29]. As a result, we used 4140 PPTX (6.56 GB), 5451 DOCX (2.22 GB) and 7124 XLSX (443 MB) files for testing and evaluation activities. Note that the testing work is executed on a desktop PC equipped with an Intel i5-4460 processor (3.2-GHz), 8 GB RAM, 256GB SSD and Microsoft Windows 8.1 (64-bits) operating system. For your guidance, the current version of SSDOC uses only a single core for the execution.

For measuring the average time cost of the prototype tool, we need to create some base user queries for finding documents having similar layouts because a core operation for retrieving queries depends mainly upon the complexity of user queries. Base user queries include all possible query types (at least one for each type), and the average cost in time for running 10 times is measured.

Table 3 displays the time consumed for each operation: extracting of layout features including parsing OOXML formats and retrieving user queries including S-value (similarity value) calculation.

*Table 3. Average cost in time of SSDOC (mm:ss)*

| OP | Operation | Time taken for 4140 PPTX (6.56 GB) | Time taken for 5451 DOCX (2.22 GB) | Time taken for 7124 XLSX (443 MB) |
|---|---|---|---|---|
| 1 | Extraction of layout features (+ parsing formats)[4] | 13:15 | 23:38 | 14:38 |
| 2 | Retrieving queries (+ calculating similarity values)[5] | 05:31 | 00:35 | 12:36 |
| 3 | Misc. (file type detection, event handling, etc.) | 00:34 | 01:25 | 00:34 |
| + | Total time taken | 19:20 | 25:38 | 27:48 |

As shown in Table 3, the most time consuming part is to extract layout features including decompressing the container structure and interpreting multiple XML files which have complex relationships with each other[6]. In order to improve this result, OP-1 could be replaced with an efficient database system having already extracted layout features. This work is one of future plans for SSDOC.

In addition, we perform an experiment with for verifying and evaluating the visual layout retrieval framework proposed here. First of all, it is necessary to determine thresholds of S-value according to the complexity of user queries in order to verify the effectiveness of the concept of treating the similarity. With determined S-value thresholds, we demonstrate the usefulness of the proposed framework through comparing results of keyword search with one of visual layout search from the viewpoint of digital forensics. This experiment is dicussed in the next section with a more detailed processes and results.

[3] Govdocs1, http://digitalcorpora.org/corpora/govdocs
[4] This part depends upon the complexity of body contents.
[5] This part depends upon the complexity of user queries and layout features extracted from the current target page (slide or sheet).
[6] The average speed of OP1 (8.45 MB/s for PPTX, 1.61 MB/s for DOCX, 0.51 MB/s for XLSX) shows that the speed of processing wordprocessing and spreadsheet files is slower than the processing speed of presentation files. This is because layout entities related to body text and cells stored in wordprocessing and spreadsheet formats are more complicated than the presentation format. (Refer to Appendix B)

# 6 Experimental Study

## 6.1 Overview

The purpose of this experimental study is to verifty the effectiveness of the proposed framework using a public dataset. In detail, this experiment shows that there is a possibility of missing potentially relevant documents when searching for files using only specific keywords.

This experiment consists of two sub-experiments: (1) determining the S-value (similarity-value) thresholds according to the complexity of user queries, (2) comparison of the traditional keyword search and the visual layout search proposed here. Note that although we performe this experiment using presentation files only, it is also possible to get similar results with wordprocessing and spreadsheet files.

## 6.2 Setup

### 6.2.1 Experimental Dataset

This experiment utilizes 4140 PPTX files converted from the GOVDOCS coupus (see Section 5.3). This dataset is suitable for our experiment because the files were collected from web servers in the .gov domain. That is, if some files were downloaded from the same web server, we may well expect that there is a chance of the existence of document files having similar layouts. For the experiment, we assume that there is no available metadata that can be utilized for filtering and classifying document files.

### 6.2.2 Study Volunteers

20 volunteers (including undergraduate and graduate students, academic researchers, and digital forensic examiners) participated in this study. They performed an experiment for determining S-value thresholds with 500 files randomly selected from the dataset. In addition, they also used the whole dataset for comparing results between keyword search and layout search in order to verifying and evaluating the visual layout retrieval framework.

## 6.3 Results and Discussion

### 6.3.1 Determining the S-value Threshold

Before achieving the purpose of this experimental study, it was necessary to determine thresholds of S-value according to the user-controlled layout queries. The S-value means the similarity level between features extracted from a target file/page (slide or sheet) and user-controlled layout queries. For that, volunteers utilized SSDOC with various layout queries defined by each of them, and analyzed the results between S-values calculated by the prototype tool and levels of feeling the similarity with the naked eye.

Figure 8 shows the S-value change depending on the number of layout queries. As shown in the graph, when users applied more numbers of queries, they thought that the target data were similar to the queries at lower S-values. In case of that the number of queries is more than 10, S-value thresholds tend to remain constant at about 0.72. Note that the S-value threshold, of course, can be adjusted high or low depending on the users' need for filtering retrieval results. In this experiment, we utilize values shown in Figure 8 to calculate the precision and recall in Section 6.3.3.

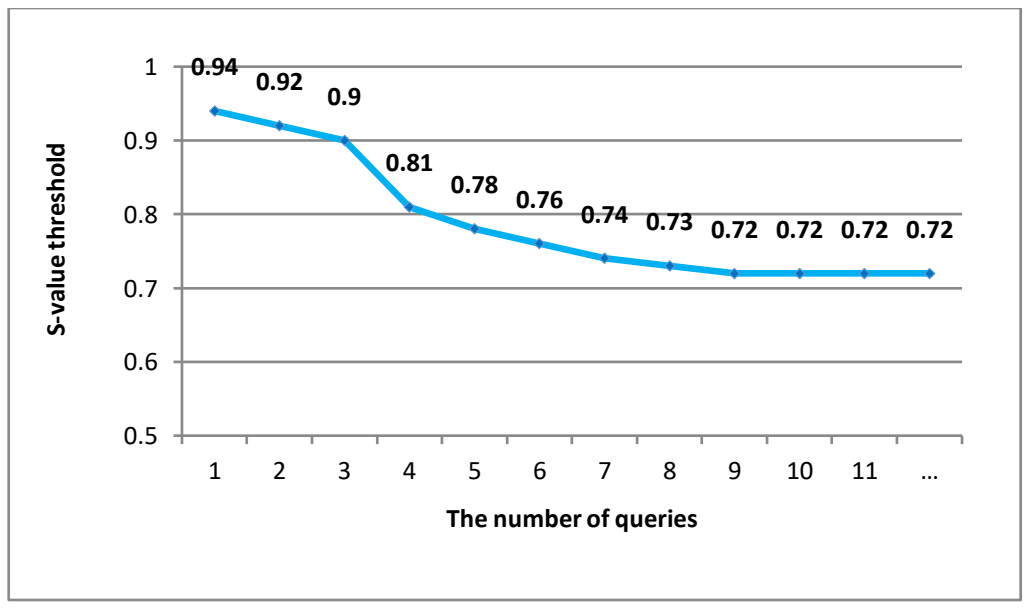

*Figure 6. S-value thresholds depending on the number of queries*

### 6.3.2 Comparison of Keyword Search and Layout Search

Volunteers first classified PPTX files having similar layouts into several groups. In this step, they considered that files in a group are relevant to each other although it has totally different contents. Among groups classified by volunteers, we represent only results of the experiment with three groups[7] to effectively explain the meaning of the proposed approach.

After grouping files, volunteers randomly selected query files (028001.pptx, 021585.pptx and 018136.pptx) from each group. The following steps were performed on each group: (1) selecting at least five words from body text of a query file that can be part of the subject, (2) performing the keyword search on the dataset using words selected in the previous step, (3) reporting results of the keyword search, (4) selecting two pages (slides) of a query file for generating layout queries, (5) performing the layout search on the dataset using queries generated in the previous step, (6) reporting results of the layout search filtered by S-value thresholds, and finally (7) comparing results between the keyword search and the layout search. For your guidance, classified presentation files and generated queries (XML files) for each layout search are included in downloadable data[3].

Table 4 summarizes processes and results of our attempts with three different groups. The first column shows each query file selected for each experiment, and the second and third columns represent words and layout queries for the keyword and layout search respectively. Diagrams in the fourth column of the table illustrate that the similar layout search proposed here allows us to find additional files which are not included in results of the keyword search.

*Table 4. Results of the comparison of keyword and layout Search*

| Query file | Words from query file | Layout queries from query file[8] | Results of keyword and layout search | |
|---|---|---|---|---|
| 028001 | Cheyenne mountain<br>Colorado springs<br>Al Pocock<br>Program philosophy<br>self-advocacy<br>student mentoring<br>ADHD | 1st slide: SWH, 3TB<br>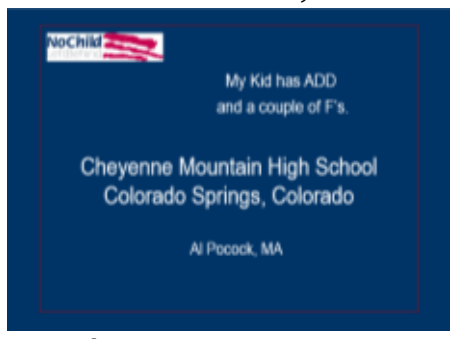<br>6th slide: SWH, 2TB<br>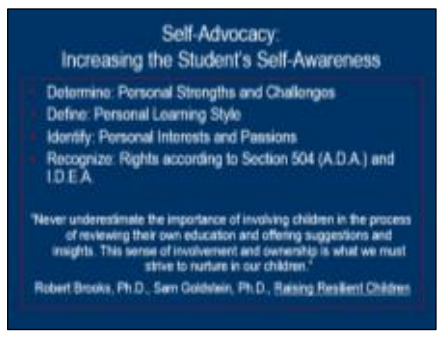 | 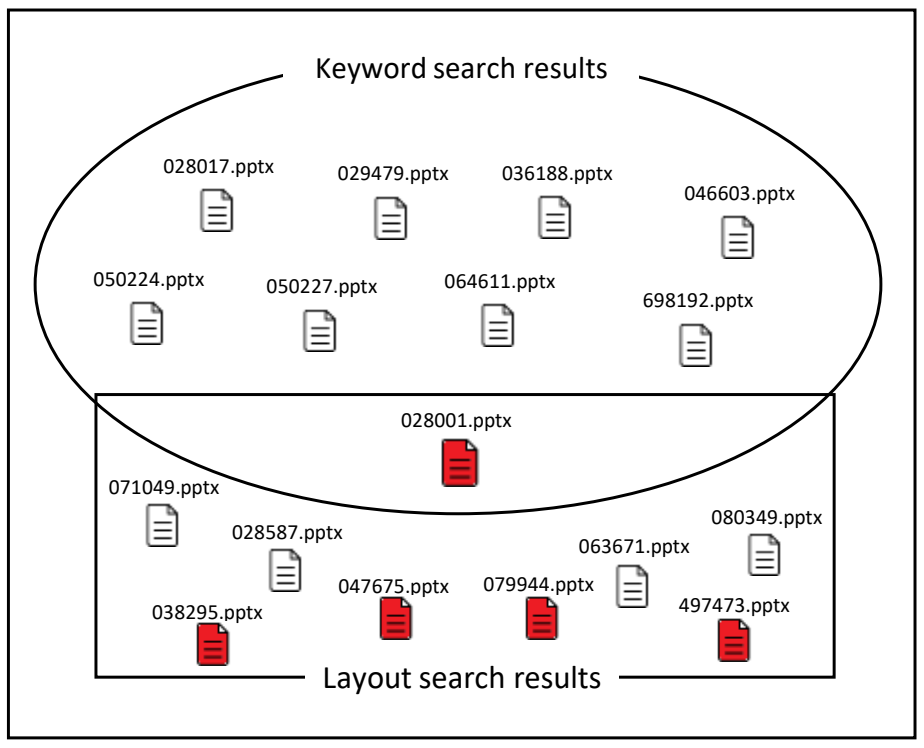 | Keyword search<br>Total: 12<br>Relevant[9]: 1<br>Others: 11<br><br>Layout search<br>Total: 9<br>Relevant: 5<br>Others: 4 |
| 021585 | HHS-348<br>COMP Time<br>Travel Issues<br>Hotel Reservations<br>NFT<br>Diane<br>osophs@ | 1st slide: SWH, 2TB, 1IMG<br>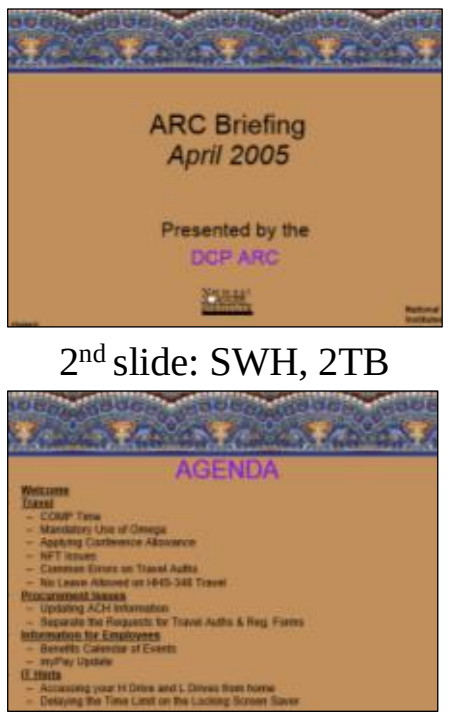<br>2nd slide: SWH, 2TB | 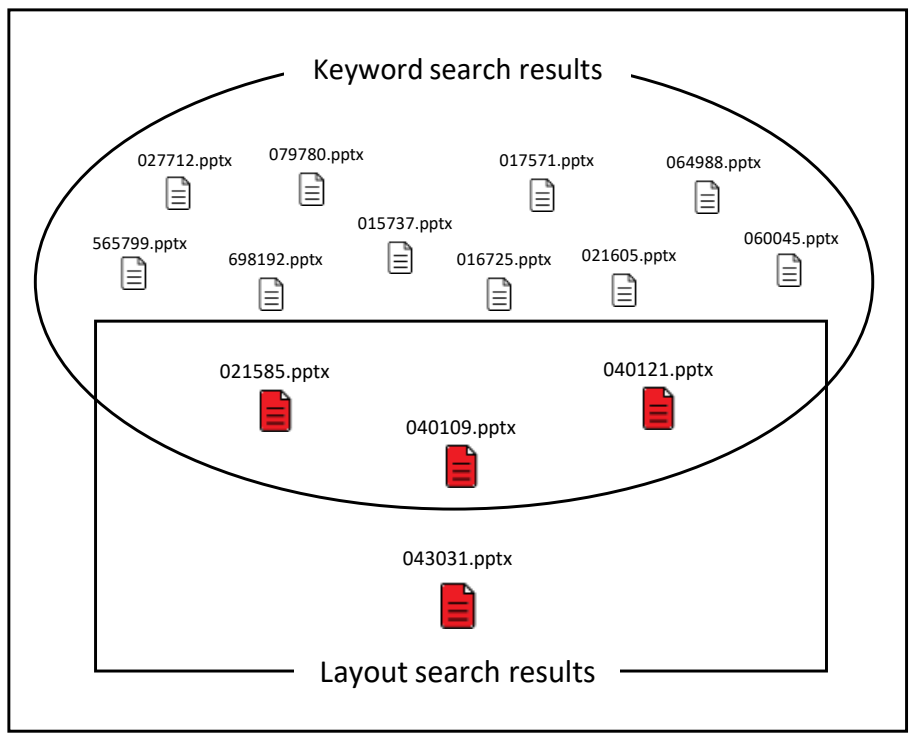 | Keyword search<br>Total: 47<br>Relevant: 3<br>Others: 44<br><br>Layout search<br>Total: 4<br>Relevant: 4<br>Others: 0 |

[7] The following is a list of files of each group:
- G1: 028001, 038295, 047675, 079944, 497473
- G2: 021585, 040109, 040121, 043031
- G3: 018136, 037886, 041312, 049417, 056160, 060047, 062953, 062956, 079951, 079955, 719239, 240013

Note that files were named by the GOVDOCS corpus, and their extensions (.pptx) were omitted.

[8] SWH (Slide Width and Height), TB (TextBox) and IMG (Image) (See Appendix B)

[9] 'Relevant' means that identified files are in this group.

<table>
<tr><td rowspan="6">018136</td><td>Proton</td><td rowspan="6">1st slide: SWH, 2TB<br>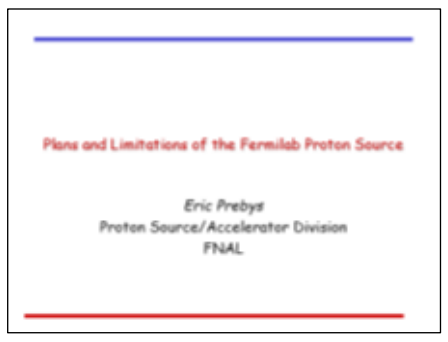<br><br>8th slide: SWH, 4TB<br>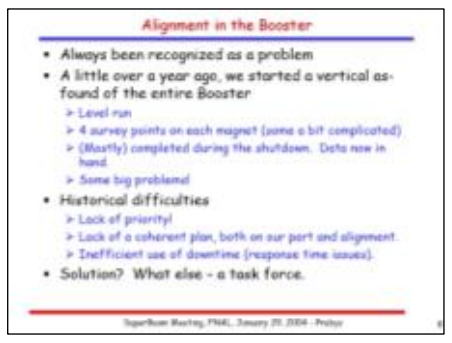<br></td><td rowspan="6">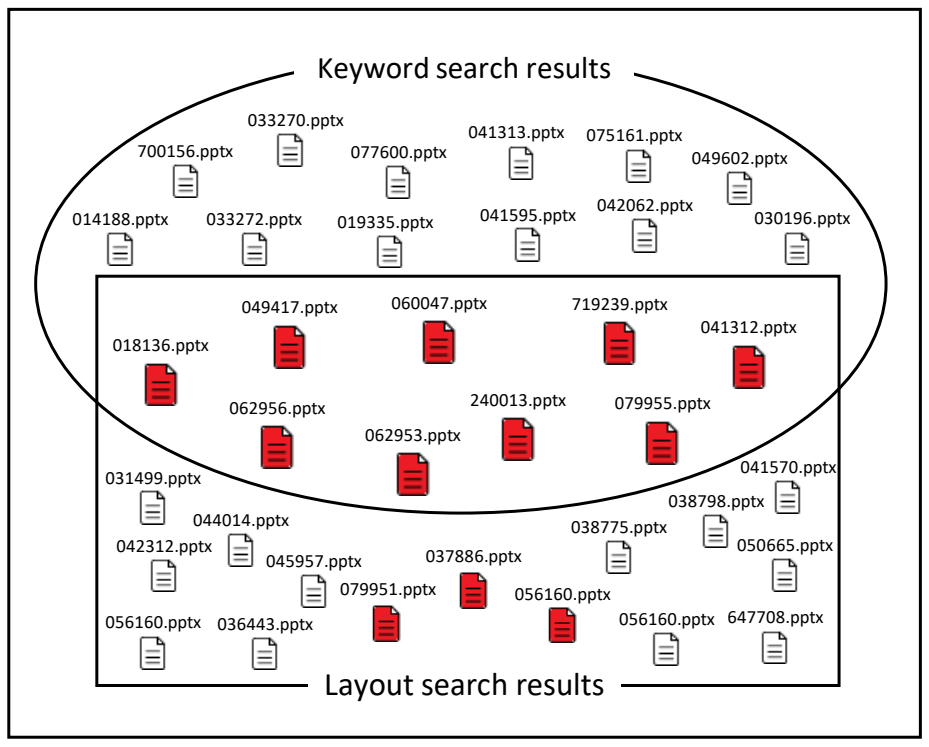<br></td><td rowspan="6"><u>Keyword search</u><br>Total: 329<br>Relevant: 9<br>Others: 320<br><br><u>Layout search</u><br>Total: 24<br>Relevant: 12<br>Others: 12</td></tr>
<tr><td>Eric Prebys</td></tr>
<tr><td>Accelerator Division</td></tr>
<tr><td>FNAL</td></tr>
<tr><td>SuperBeam</td></tr>
<tr><td>Energy loss</td></tr>
</table>

### 6.3.3 Precision and Recall of Layout Search

Detailed results of the layout search are summarized in Table 5. Each row contains the precision and recall on each set of user layout queries. As shown in the table, the proposed method performed a quite effective retrieval for finding potentially relevant document files. Interestingly, results from the third group (G3) represent a situation where multiple sets of user queries are required for achieving more accurate retrieval results. This is because some files were not found with the S-value threshold 0.90 when the 1st slide of 018136 file was used as a set of queries. However, when five layout entities from the 8th slide of 018136 file were used as a set of user queries, we could find all document files in G3.

*Table 5. Precision and recall details*

| **Group** | **Layout query** (refer to the 2nd column in Table 4) | **S-value threshold** | **The number of similar slides that are searched as similar slides (A)** | **The number of dissimilar slides that are searched as similar slides (B)** | **The number of similar slides that are searched as dissimilar slides (C)** | **Precision A/(A+B)** | **Recall A/(A+C)** |
|---|---|---|---|---|---|---|---|
| G1 | 028001 (1st slide) query count: 4 | 0.81 | 5 | 1 | 0 | 0.83 | 1 |
| | 028001 (6th slide) query count: 3 | 0.90 | 5 | 3 | 0 | 0.63 | 1 |
| G2 | 021585 (1st slide) query count: 4 | 0.81 | 4 | 0 | 0 | 1 | 1 |
| | 021585 (2nd slide) query count: 3 | 0.90 | 4 | 0 | 0 | 1 | 1 |
| G3 | 018136 (1st slide) query count: 3 | 0.90 | 9 | 5 | 3 | 0.64 | 0.75 |
| | 018136 (8th slide) query count: 5 | 0.78 | 12 | 7 | 0 | 0.63 | 1 |

It is important to note that the precision and recall in this study mainly depends on groups classified by humans and S-value thresholds for filtering retrieval results. There are also other important factors regarding how many sets of user queries are used and how to build user queries. In particular, S-value thresholds used here are not fixed values because it will vary depending on what kind of dataset is used or who determines the value, and so our framework allows users to adjust the S-value threshold according to their needs as mentioned in Section 6.3.1. Therefore, we only explained the precision and recall of results on three different groups (G1, G2 and G3) with pre-defined S-value thresholds as an example.

Through above experiments, we demonstrated that our approach on retrieving similar document files is useful and helpful for addressing possible situations like the one described in section 2.

## 7 Conclusion and Future Works

Currently, search techniques focusing on data similarity are meaningful, because the number of digital devices requiring investigation are increasing along with the amount of digital documents. Thus, similar data searches become more important from the viewpoint of information retrieval and digital forensics. Existing studies associated with data similarity have mainly focused on byte-stream-based, structure-based, and content-based similarities. These studies have been useful for digital forensic purposes, however, their methods and techniques are not suitable for situations where more efficient electronic document retrieval based on layout similarity is required for a specific investigative purpose as shown in a case described in Section 2.

For these reasons, this study proposed a new framework for retrieving digital document files containing similar visual layouts based on the characteristics of each file format. Additionally, we designed and developed SSDOC that is a prototype tool capable of searching similar Microsoft OOXML files based on the proposed framework. We also performed an experiment for verifying and evaluating the prototype tool using a public dataset. This experiment verified that the tool can successfully find potentially relevant document files having similar layouts by using user-controlled queries. The experimental result also suggested that the similar layout search is useful for digital forensic activities if it can be utilized appropriately with the traditional keyword search. Therefore, if investigators utilize the method proposed here, they will be able to perform their work more accurately and efficiently.

In the future, we will extend our research to the retrieval of drawing files having similar drawing objects. Moreover, the current prototype version of SSDOC will also be enhanced with an efficient database system to enable processing and indexing large-scale datasets.

*Appendix A. Four types of AM (Approximate Matching) algorithm*

| Type | Method to calculate S-value (similarity value) | S-value range |
|---|---|---|
| AM-1 | $\text{S_value} = \frac{\sum_{i=1}^{k} s_value_i}{k}$, k = query count | [0, 1] |
| AM-2 | S_value = 1.0, if definitely match<br>S_value = 0.5, if type match, but dimension is not match<br>S_value = 0.0, if not match | 0 or 0.5 or 1 |
| AM-3 | 1) <u>Compute the distance range</u><br>distance_range = max(abs($k - min_distance$), abs(k $-$ $max_distance$))<br>, where $k$ is a user query<br><br>2) <u>Exponential distribution</u><br>$\text{S_value} = 4 * \lambda e^{-\lambda x}$ , $x = \text{distance_range}$, $x \geq 0$, $\lambda = \text{maximum value } (0.25)$ | [0, 1] |
| AM-4 | 1) <u>Compute the maximum distance</u><br>$\text{max_distance} = \sqrt{(x_2 - x_1)^2 + (y_2 - y_1)^2}$<br>, where a query coordinate $C_q$ is $(x_1, y_1)$ and the farthest coordinate from $C_q$ is $(x_2, y_2)$<br><br>2) <u>Exponential distribution</u><br>$\text{S_value} = 4 * \lambda e^{-\lambda d}$ , $d = \text{max_distance}$, $d \geq 0$, $\lambda = \text{maximum value } (0.25)$ | [0, 1] |

*Appendix B. Methods to calculate the similarity-value depending on user query types for MS OOXML files*

| File format | Type (Abbreviation) | Subtype | Example | Description | Method to calculate S-value (max: max distance, min: min distance) |
|---|---|---|---|---|---|
| DOCX | Page layout (PL) | Height | 27.97 | Centimeters or Inches | AM-3 max = 55.87(cm) |
| | | Width | 21.59 | Centimeters or Inches | |
| | | Column | 2 | Integer | |
| | | Column margin | 0.5 | Centimeters or Inches | |
| | | Upper margin | 2.54 | Centimeters or Inches | |
| | | Right margin | 3.17 | Centimeters or Inches | |
| | | Lower margin | 2.54 | Centimeters or Inches | |
| | | Left margin | 3.17 | Centimeters or Inches | |
| | | Header margin | 1.27 | Centimeters or Inches | |
| | | Footer margin | 1.27 | Centimeters or Inches | |
| | | Gutter margin | 2.51 | Centimeters or Inches | |
| | Text (TXT) | Font size | 12, 11, 12 | Retrieval using each subtype or multiple subtypes with comma-separated values | AM-1 |
| | | Font color | 000000, 0070C0, 000000 | | |
| | | Font name | Times New Roman, Arial, Calibri | | |
| | Footnote (FNT) | Font size | 20 | Integer | AM-1 |
| | | Font color | FF0000 | Hex. color codes (RGB) | |
| | | Font name | Candara | Case-insensitive | |
| | Header (HDR) | Font size | 10 | Integer | AM-1 |
| | | Font color | 000000 | Hex. color codes (RGB) | |
| | | Font name | Verdana | Case-insensitive | |
| | Footer (FTR) | Font size | 10 | Integer | AM-1 |
| | | Font color | 165189 | Hex. color codes (RGB) | |
| | | Font name | Verdana | Case-insensitive | |
| | Image (IMG) | Height | 18.99 | Centimeters or Inches | AM-3 max = width or height of a page |
| | | Width | 14.18 | Centimeters or Inches | |
| | Table (TBL) | Row | 3 | Integer | EM |
| | | Column | 4 | Integer | EM |
| | | Font size | 12,10 | Retrieval using each subtype or multiple subtypes with comma-separated values | AM-1 |
| | | Font color | 000000,000000 | | |
| | | Font name | Arial, Consolas | | |
| PPTX | Slide width & height (SWH) | Height | 25.4 | Centimeters or Inches | AM-3 max = 142.24(cm), min = 2.54(cm) |
| | | Width | 19.05 | Centimeters or Inches | |
| | Textbox (TB) | Coordinate X | 1.06 | Centimeters or Inches | AM-4 |
| | | Coordinate Y | 4.02 | Centimeters or Inches | |
| | | Height | 12.90 | Centimeters or Inches | AM-3 max = width or height of a slide |
| | | Width | 23.28 | Centimeters or Inches | |
| | | Font name | Times New Roman | Case-insensitive | AM-1 |
| | | Font color | 000000, FFFFFF | Hex. color codes (RGB) | AM-1 |
| | Image (IMG) | Coordinate X | 3.25 | Centimeters or Inches | AM-4 |
| | | Coordinate Y | 4.55 | Centimeters or Inches | |
| | | Height | 2.66 | Centimeters or Inches | AM-3 max = width or height of a slide |
| | | Width | 7.22 | Centimeters or Inches | |
| | Table (TBL) | Coordinate X | 5.26 | Centimeters or Inches | AM-4 |
| | | Coordinate Y | 1.26 | Centimeters or Inches | |
| | | Height | 2.49 | Centimeters or Inches | AM-3 max = width or height of a slide |
| | | Width | 9.15 | Centimeters or Inches | |
| | | Row | 3 | Integer | EM |
| | | Column | 2 | Integer | EM |
| | | Font name | Times New Roman | Case-insensitive | AM-1 |
| | | Font color | FF0000 | Hex. color codes (RGB) | AM-1 |
| | Shape (SH) | Shape type | LeftRightArrow | Case-insensitive | EM |
| | | Coordinate X | 10.10 | Centimeters or Inches | AM-4 |
| | | Coordinate Y | 15.32 | Centimeters or Inches | |
| | | Height | 5.12 | Centimeters or Inches | AM-3 max = width or height of a slide |
| | | Width | 6.85 | Centimeters or Inches | |
| XLSX1 | Zoom scale (ZS) | - | 85 | Integer | AM-3 max = 142.24(cm), min = 10(cm) |
| | Cell | Font size (FTS) | 12, 11, 11 | Retrieval using each subtype or multiple subtypes with comma-separated values | AM-1 |
| | | Font name (FTN) | CG Times (WN), Calibri, Calibri | | |
| | | Fill pattern (FIP) | None, yellow, yellow | | |
| | | Fill color (FIC) | gray0625 | | |
| | | Border (BRD) | double, dotted, thin, none | | |
| | Image | Image position 'from' cell (IMGF) | R1C5 | R1C1 reference style | AM-4 |
| | | Image position 'to' cell (IMGT) | R5C10 | R1C1 reference style | AM-4 |
| | Chart | Chart type (CHTY) | barChart | Case-insensitive | AM-2 |
| | | Chart position 'from' cell (CHTF) | R8C9 | R1C1 reference style | AM-4 |
| | | Chart position 'to' cell (CHTT) | R21C15 | R1C1 reference style | AM-4 |

*Appendix C. Sample pages of files used in Section 6.3.2*

| Query file | Sample pages of other files in three groups | | | |
|---|---|---|---|---|
| 028001.pptx (p 1)<br>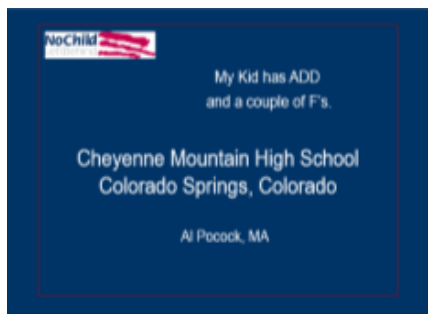 | 497473.pptx (p 1)<br>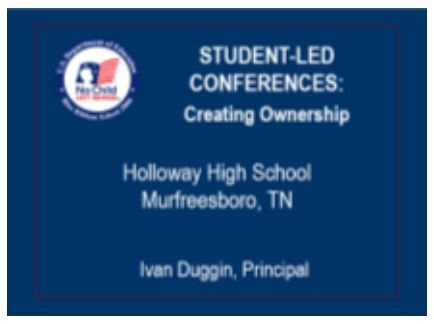 | 038295.pptx (p 1)<br>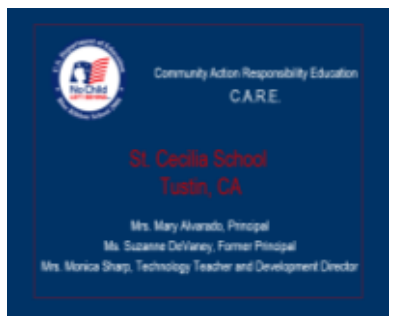 | 047675.pptx (p 1)<br>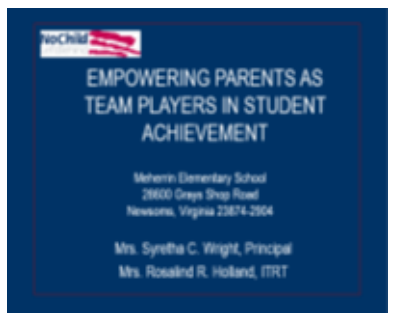 | 079944.pptx (p 1)<br>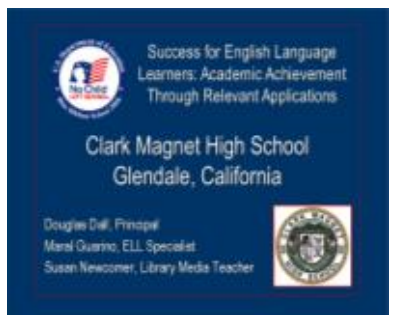 |
| 028001.pptx (p 6)<br>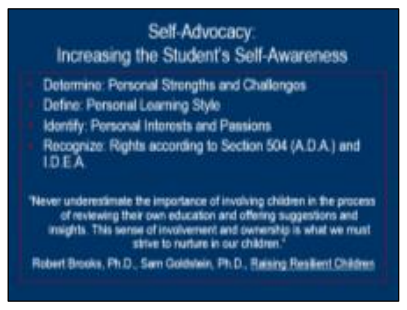 | 497473.pptx (p 3)<br>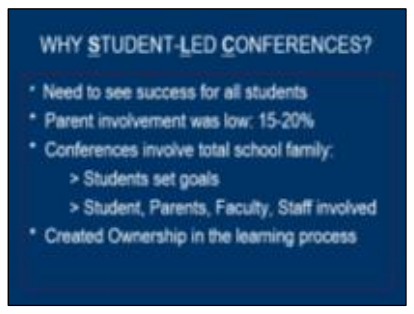 | 038295.pptx (p 8)<br>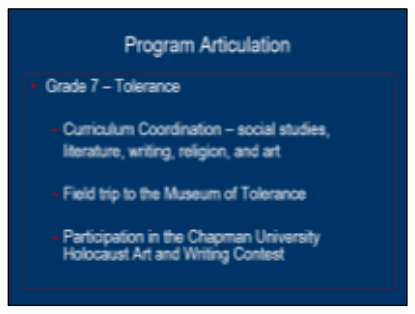 | 047675.pptx (p 3)<br>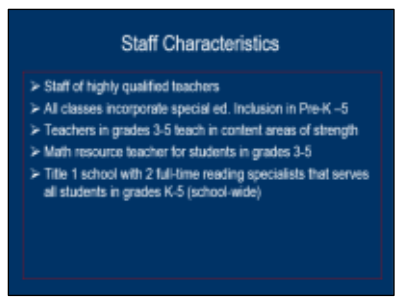 | 079944.pptx (p 6)<br>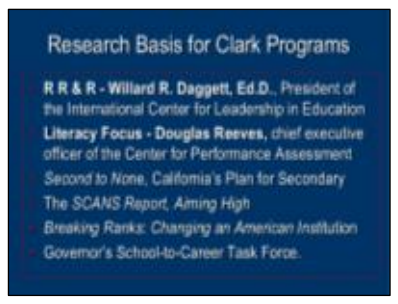 |
| 021585.pptx (p 1)<br>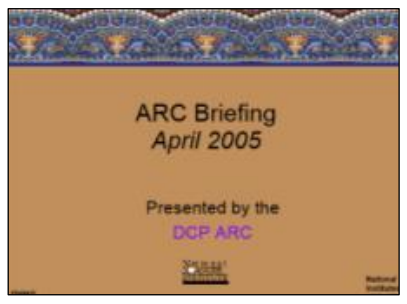 | 040109.pptx (p 1)<br>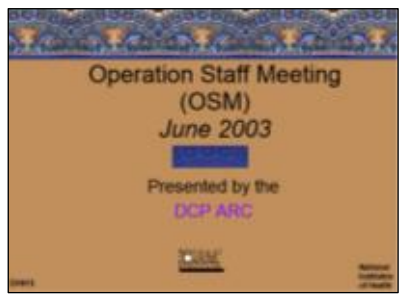 | 040121.pptx (p 1)<br>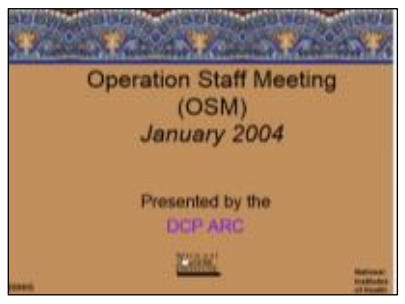 | 043031.pptx (p 1)<br>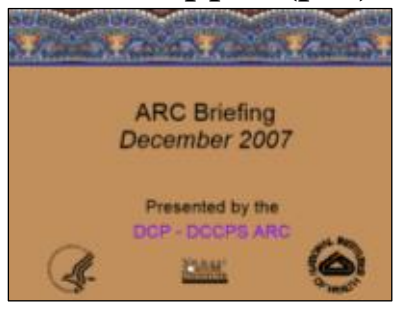 | - |
| 021585.pptx (p 2)<br>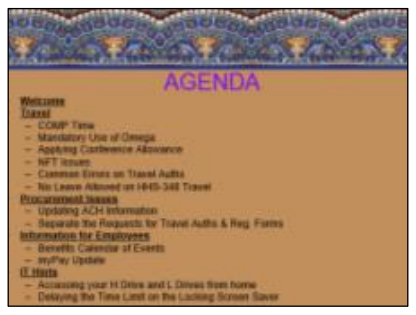 | 040109.pptx (p 6)<br>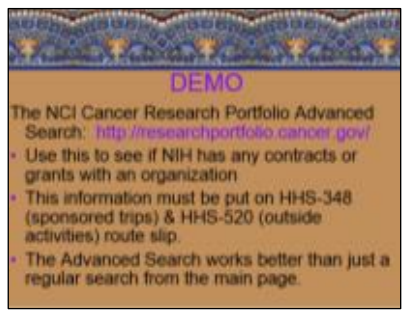 | 040121.pptx (p 10)<br>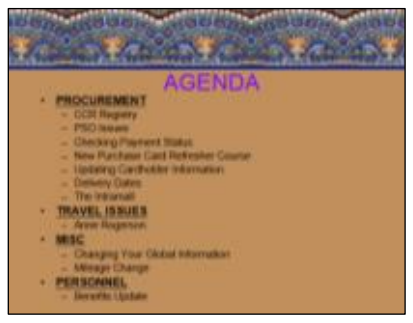 | 043031.pptx (p 2)<br>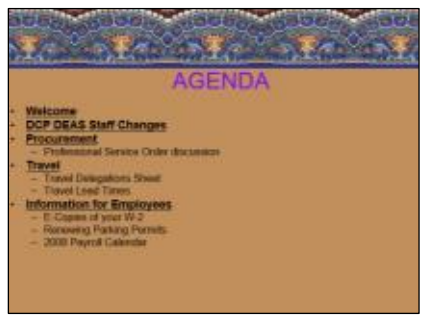 | - |
| 018136.pptx (p 1)<br>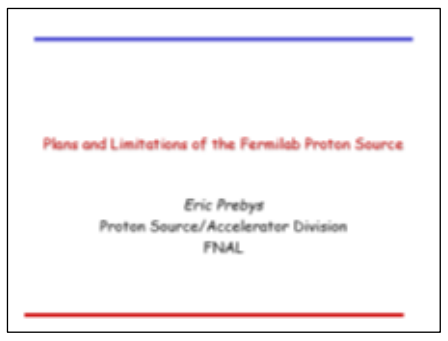 | 062953.pptx (p 1)<br>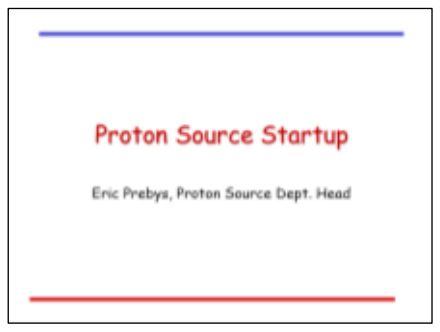 | 079955.pptx (p 1)<br>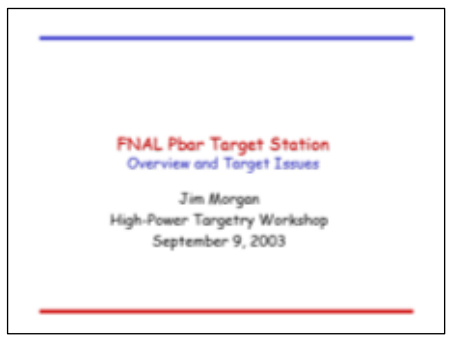 | 041312.pptx (p 1)<br>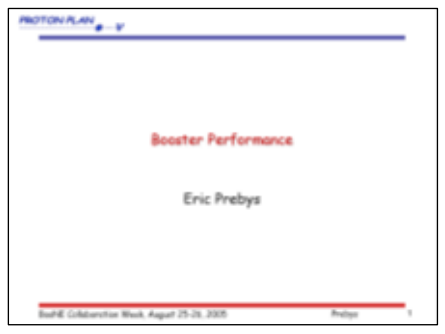 | 056160.pptx (p 1)<br>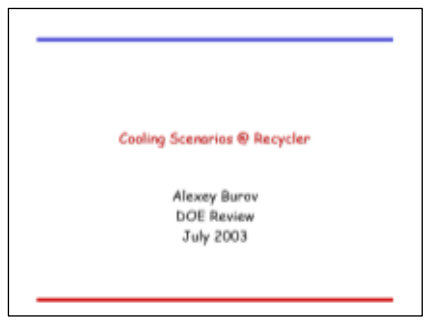 |
| 018136.pptx (p 8)<br>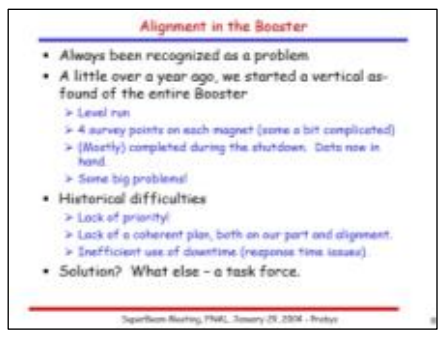 | 049417.pptx (p 2)<br>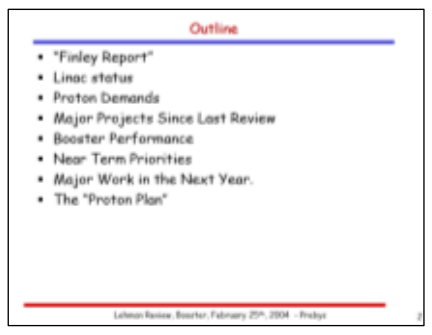 | 079951.pptx (p 3)<br>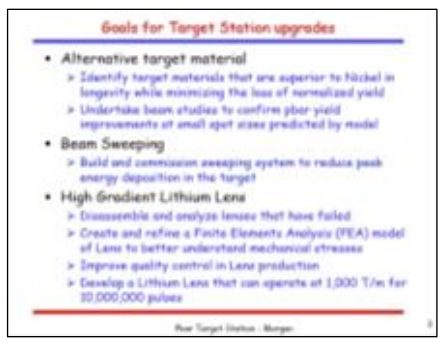 | 240013.pptx (p 4)<br>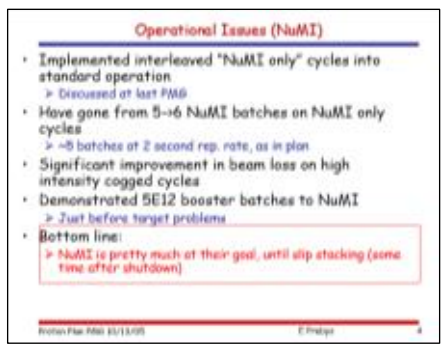 | 056160.pptx (p 2)<br>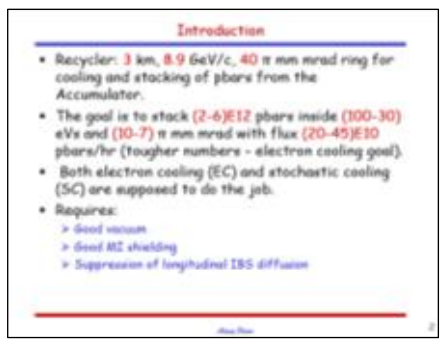 |